\documentclass[12pt]{article}
\usepackage{amssymb}
\begin{document}

%************************** Text Begins here ******************************

%  Greek letters

\def\a{\alpha}
\def\b{\beta}
\def\d{\delta}
\def\e{\epsilon}
\def\g{\gamma}
\def\h{\mathfrak{h}}
\def\k{\kappa}
\def\l{\lambda}
\def\o{\omega}
\def\p{\wp}
\def\r{\rho}
\def\t{\tau}
\def\s{\sigma}
\def\z{\zeta}
\def\x{\xi}
 \def\A{{\cal{A}}}
 \def\B{{\cal{B}}}
 \def\C{{\cal{C}}}
 \def\D{{\cal{D}}}
\def\G{\Gamma}
\def\K{{\cal{K}}}
\def\O{\Omega}
\def\R{\bar{R}}
\def\T{{\cal{T}}}
\def\L{\Lambda}
\def\f{E_{\tau,\eta}(sl_2)}
\def\E{E_{\tau,\eta}(sl_n)}
\def\Zb{\mathbb{Z}}
\def\Cb{\mathbb{C}}

\def\R{\overline{R}}
% Shorthands for \begin{equation} and the like

\def\beq{\begin{equation}}
\def\eeq{\end{equation}}
\def\bea{\begin{eqnarray}}
\def\eea{\end{eqnarray}}
\def\ba{\begin{array}}
\def\ea{\end{array}}
\def\no{\nonumber}
\def\le{\langle}
\def\re{\rangle}
\def\lt{\left}
\def\rt{\right}

\newtheorem{Theorem}{Theorem}
\newtheorem{Definition}{Definition}
\newtheorem{Proposition}{Proposition}
\newtheorem{Lemma}{Lemma}
\newtheorem{Corollary}{Corollary}
\newcommand{\proof}[1]{{\bf Proof. }
        #1\begin{flushright}$\Box$\end{flushright}}

\baselineskip=20pt

%%%%%%%%%%%%%%%%%%%%%%%%%%%%%%%%%%%%%%%%%%%%%%%%%%%%%%%%%%%%
%                                                          %
%  Title page                                              %
%                                                          %
%%%%%%%%%%%%%%%%%%%%%%%%%%%%%%%%%%%%%%%%%%%%%%%%%%%%%%%%%%%%
\newfont{\elevenmib}{cmmib10 scaled\magstep1}
\newcommand{\preprint}{
   \begin{flushleft}
     %\elevenmib Yukawa\, Institute\, Kyoto\\
   \end{flushleft}\vspace{-1.3cm}
   \begin{flushright}\normalsize
  % \sf  YITP-03-53\\
     {\tt hep-th/0411048} \\ June 2004
   \end{flushright}}
\newcommand{\Title}[1]{{\baselineskip=26pt
   \begin{center} \Large \bf #1 \\ \ \\ \end{center}}}
\newcommand{\Author}{\begin{center}
   \large \bf
Wen-Li Yang$,{}^{a,b}$
 ~ Yao-Zhong Zhang ${}^b$ and~Mark D. Gould${}^b$\end{center}}
\newcommand{\Address}{\begin{center}

     ${}^a$ Institute of Modern Physics, Northwest University,
     Xian 710069, P.R. China\\
     ${}^b$ Department of Mathematics, University of Queensland, Brisbane, QLD 4072, Australia
   \end{center}}
\newcommand{\Accepted}[1]{\begin{center}
   {\large \sf #1}\\ \vspace{1mm}{\small \sf Accepted for Publication}
   \end{center}}

\preprint
\thispagestyle{empty}
\bigskip\bigskip\bigskip

\Title{Exact solution of the XXZ  Gaudin
      model with generic open boundaries } \Author

\Address
\vspace{1cm}

\begin{abstract}
The XXZ Gaudin model with  {\it generic\/} integerable boundaries
specified by generic {\it non-diagonal\/} K-matrices is studied.
The commuting families of Gaudin operators are diagonalized by the
algebraic Bethe ansatz method. The eigenvalues and the
corresponding Bethe ansatz equations are obtained.

\vspace{1truecm} \noindent {\it PACS:} 03.65.Fd; 04.20.Jb;
05.30.-d; 75.10.Jm

\end{abstract}
\newpage
%%%%%%%%%%%%%%%%%%%%%%%%%%%%%%%%%%%%%%%%%%%%%%%%%%%%%%%%%%%%%%%
%                                                             %
%  1. Introduction                                            %
%                                                             %
%%%%%%%%%%%%%%%%%%%%%%%%%%%%%%%%%%%%%%%%%%%%%%%%%%%%%%%%%%%%%%%
\section{Introduction}
\label{intro} \setcounter{equation}{0}

Gaudin type models \cite{Gau76} have played an essential  role in
the study of quantum systems appeared in many physical fields such
as the BCS theory \cite{Bar57} of small metallic grains
\cite{Cam97,Ami01, Del01}, nuclear physics theory
\cite{Dea03,Duk04} and QCD theory \cite{Iac94, Ris00}. They also
provide  a testing ground for ideas such as the functional Bethe
ansatz and general procedure of separation of variables
\cite{Skl89,Skl96,Brz94} and the construction of integral
representations of the solutions to the Knizhnik-Zamolodchikov
(KZ) equation \cite{Bab93,Has94,Fei94,Hik95,Gou02}.

The original Gaudin's magnet Hamiltonians (or Gaudin operators)
can be rewritten in terms of the quasi-classical expansion of the
transfer matrices ({\it row-to-row transfer matrices\/}) of
inhomogeneous spin chains with periodic  boundary condition
\cite{Hik92}. Since the elegant work of Sklyanin \cite{Skl88}, the
powerful Quantum Inverse Scattering Method (QISM) has been applied
to various integrable models with non-trivial boundary conditions,
which are specified by K-matrices satisfying the reflection
equation  and its dual \cite{Che84}. The quasi-classical expansion
of the corresponding transfer matrices ({\it double-row transfer
matrices\/}) produces {\it generalized\/} Gaudin Hamiltonians with
boundaries \cite{Hik95, Lor02}. In particular, twisted boundary
conditions and open boundary conditions associated with {\it
diagonal\/} K-matrices give rise to Gaudin magnets in non-uniform
local magnetic fields \cite{Hik95} and interacting electron pairs
with certain non-uniform long-range coupling strengths
\cite{Ami01,Duk01,Zho02,Del02,Lor02}.

In this paper, we  study the XXZ type Gaudin magnets with most
generic boundary conditions specified by the generic {\it
non-diagonal\/} K-matrices given in \cite{Veg93,Gho94}. In section
3, we construct the generalized Gaudin operators associated with
the generic boundary K-matrices. The commutativity of these
operators follows from the standard procedure
\cite{Hik92,Hik95,Lor02} specializing to the inhomogeneous
spin-$\frac{1}{2}$ XXZ open chain, thus ensuring the integrability
of the Gaudin magnets. In section 4, we diagonalize the Gaudin
operators simultaneously by means of the algebraic Bethe ansatz
method. This constitutes the main new result in this paper. The
diagonalization is achieved by means of the technique of the
``vertex-face" transformation. In section 5, we conclude this
paper by offering some discussions.

%%%%%%%%%%%%%%%%%%%%%%%%%%%%%%%%%%%%%%%%%%%%%%%%%%%%%%%%%%%%%%%
%                                                             %
%  2. Preliminaries: the inhomogeneous spin-$\frac{1}{2}$     %
%                    XXZ open chain                           %
%                                                             %
%                                                             %
%%%%%%%%%%%%%%%%%%%%%%%%%%%%%%%%%%%%%%%%%%%%%%%%%%%%%%%%%%%%%%%

\section{ Preliminaries: the inhomogeneous spin-$\frac{1}{2}$ XXZ open chain}
\label{XXZ} \setcounter{equation}{0}

Throughout, $V$ denotes a two-dimensional linear space and
$\s^{\pm},\,\s^z$ are the usual Pauli matrices which realize the
spin-$\frac{1}{2}$ representation of the Lie algebra $sl(2)$ on
$V$. The spin-$\frac{1}{2}$ XXZ chain can be constructed from the
well-known six-vertex model R-matrix $\R(u)\in {\rm End}(V\otimes
V)$ \cite{Kor93} given by \bea
\R(u)=\lt(\begin{array}{llll}a(u)&&&\\&b(u)&c(u)&\\
&c(u)&b(u)&\\&&&a(u)\end{array}\rt).\label{r-matrix}\eea The
coefficient functions read \bea
&&\R^{11}_{11}(u)=\R^{22}_{22}(u)=a(u)=1,\no\\
&&\R^{12}_{12}(u)=\R^{21}_{21}(u)=b(u)=\frac{\sin
u}{\sin(u+\eta)},\no\\
&&\R^{12}_{21}(u)=\R^{21}_{12}(u)=c(u)=\frac{\sin\eta}{\sin(u+\eta)}.
\label{r-matrix-3}\eea Here $u$ is the spectrum parameter and
$\eta$ is the so-called crossing parameter. The R-matrix satisfies
the quantum Yang-Baxter equation (QYBE), \bea
R_{12}(u_1-u_2)R_{13}(u_1-u_3)R_{23}(u_2-u_3)
=R_{23}(u_2-u_3)R_{13}(u_1-u_3)R_{12}(u_1-u_2),\label{QYB}\eea and
the properties, \bea &&\hspace{-1.5cm}\mbox{
Unitarity}:\,{\R}_{12}(u)\R_{21}(-u)= {\rm id},\label{Unitarity}\\
&&\hspace{-1.5cm}\mbox{
Crossing-unitarity}:\,(\R)^{t_2}_{21}(-u-2\eta)(\R)_{12}^{t_2}(u)
=\frac{\sin u\sin(u+2\eta)}{\sin(u+\eta)\sin(u+\eta)}~{\rm id},
\label{crosing-unitarity}\\
&&\hspace{-1.5cm}\mbox{ Quasi-classical
property}:\,{\R}_{12}(u)|_{\eta\rightarrow 0}= {\rm
id}.\label{quasi} \eea Here
$\bar{R}_{21}(u)=P_{12}\bar{R}_{12}(u)P_{12}$ with $P_{12}$ being
the usual permutation operator and $t_i$ denotes the transposition
in the $i$-th space. Here and below we adopt the standard
notations: for any matrix $A\in {\rm End}(V)$ , $A_j$ is an
embedding operator in the tensor space $V\otimes V\otimes\cdots$,
which acts as $A$ on the $j$-th space and as identity on the other
factor spaces; $R_{ij}(u)$ is an embedding operator of R-matrix in
the tensor space, which acts as identity on the factor spaces
except for the $i$-th and $j$-th ones.

One introduces the ``row-to-row" monodromy matrix $T(u)$, which is
an $2\times 2$ matrix with elements being operators acting on
$V^{\otimes N}$, where $N=2M$ ($M$ being a positive integer),\bea
T_0(u)=\R_{01}(u+z_1)\R_{02}(u+z_2)\cdots
\R_{0N}(u+z_N).\label{Mon-V}\eea Here $\{z_j|j=1,\cdots,N\}$ are
arbitrary free complex parameters which are usually called
inhomogeneous parameters. With the help of the QYBE (\ref{QYB}),
one can show that $T(u)$ satisfies the so-called ``RLL" relation
\bea
\R_{12}(u-v)T_1(u)T_2(v)=T_2(v)T_1(u)\R_{12}(u-v).\label{Relation1}\eea

Integrable open chain can be constructed as follows \cite{Skl88}.
Let us introduce a pair of K-matrices $K^-(u)$ and $K^+(u)$. The
former satisfies the reflection equation (RE) \cite{Che84}
 \bea &&\R_{12}(u_1-u_2)K^-_1(u_1)\R_{21}(u_1+u_2)K^-_2(u_2)\no\\
 &&~~~~~~=
K^-_2(u_2)\R_{12}(u_1+u_2)K^-_1(u_1)\R_{21}(u_1-u_2),\label{RE-V}\eea
and the latter  satisfies the dual RE \bea
&&\R_{12}(u_2-u_1)K^+_1(u_1)\R_{21}(-u_1-u_2-2\eta)K^+_2(u_2)\no\\
&&~~~~~~=
K^+_2(u_2)\R_{12}(-u_1-u_2-2\eta)K^+_1(u_1)\R_{21}(u_2-u_1).
\label{DRE-V}\eea For open spin-chains, instead of the standard
``row-to-row" monodromy matrix $T(u)$ (\ref{Mon-V}), one needs to
introduce the
 ``double-row" monodromy matrix $\mathbb{T}(u)$ \bea
\mathbb{T}(u)=T(u)K^-(u)T^{-1}(-u).\label{Mon-V-1}\eea Using
(\ref{Relation1}) and (\ref{RE-V}), one can prove that
$\mathbb{T}(u)$ satisfies
 \bea \R_{12}(u_1-u_2)\mathbb{T}_1(u_1)\R_{21}(u_1+u_2)
 \mathbb{T}_2(u_2)=
\mathbb{T}_2(u_2)\R_{12}(u_1+u_2)\mathbb{T}_1(u_1)\R_{21}(u_1-u_2).
\label{Relation-Re}\eea Then the {\it double-row transfer
matrix\/} of  the inhomogeneous spin-$\frac{1}{2}$ XXZ chain with
open boundary is given by \bea
\t(u)=tr(K^+(u)\mathbb{T}(u)).\label{trans}\eea With the help of
(\ref{QYB})-(\ref{crosing-unitarity}) and
(\ref{RE-V})-(\ref{DRE-V}), one can prove that the transfer
matrices with different spectral parameters commute with each
other \cite{Skl88}: \bea [\t(u),\t(v)]=0.\label{Com-2}\eea This
ensures the integrability of the inhomogeneous spin-$\frac{1}{2}$
XXZ chain with open boundary.

%%%%%%%%%%%%%%%%%%%%%%%%%%%%%%%%%%%%%%%%%%%%%%%%%%%%%%%%%%%%%%%
%                                                             %
%  3. XXZ Gaudin model with generic boundaries            %
%                                                             %
%                                                             %
%%%%%%%%%%%%%%%%%%%%%%%%%%%%%%%%%%%%%%%%%%%%%%%%%%%%%%%%%%%%%%%
\section{XXZ Gaudin models with generic boundaries}
 \label{Haml} \setcounter{equation}{0}
In this paper, we will consider a {\it generic\/} K-matrix
$K^{-}(u)$ which is a generic solution to the RE (\ref{RE-V})
associated the six-vertex model R-matrix  \cite{Veg93,Gho94}
\bea K^-(u)=\lt(\begin{array}{ll}k_1^1(u)&k^1_2(u)\\
k^2_1(u)&k^2_2(u)\end{array}\rt)\equiv K(u).\label{K-matrix}\eea
The coefficient functions are \bea && k^1_1(u)=
\frac{2\cos(\l_1-\l_2) -\cos(\l_1+\l_2+2\xi)e^{-2iu}}
{4\sin(\l_1+\xi+u)
\sin(\l_2+\xi+u)},\no\\
&&k^1_2(u)=\frac{-i\sin(2u)e^{-i(\l_1+\l_2)} e^{-iu}}
{2\sin(\l_1+\xi+u) \sin(\l_2+\xi+u)},\no\\
&&k^2_1(u)=\frac{i\sin(2u)e^{i(\l_1+\l_2)} e^{-iu}}
{2\sin(\l_1+\xi+u) \sin(\l_2+\xi+u)}, \no\\
&& k^2_2(u)=\frac{2\cos(\l_1-\l_2)e^{-2iu}- \cos(\l_1+\l_2+2\xi)}
{4\sin(\l_1+\xi+u)\sin(\l_2+\xi+u)}.\label{K-matrix-2-1} \eea At
the same time, we introduce  the corresponding {\it dual\/}
K-matrix $K^+(u)$ which is a generic solution to the dual
reflection equation (\ref{DRE-V}) with a particular choice of the
free boundary parameters with respect to $K^-(u)$:
\bea K^+(u)=\lt(\begin{array}{ll}{k^+}_1^1(u)&{k^+}^1_2(u)\\
{k^+}^2_1(u)&{k^+}^2_2(u)\end{array}\rt).\label{DK-matrix}\eea The
matrix elements are \bea && {k^+}^1_1(u)=
\frac{2\cos(\l_1-\l_2)e^{-i\eta}
-\cos(\l_1+\l_2+2\xi)e^{2iu+i\eta}} {4\sin(\l_1+\xi-u-\eta)
\sin(\l_2+\xi-u-\eta)},\no\\
&&{k^+}^1_2(u)=\frac{i\sin(2u+2\eta)e^{-i(\l_1+\l_2)}
e^{iu-i\eta}} {2\sin(\l_1+\xi-u-\eta) \sin(\l_2+\xi-u-\eta)},
\no\\
&&{k^+}^2_1(u)=\frac{-i\sin(2u+2\eta)e^{i(\l_1+\l_2)}
e^{iu+i\eta}}
{2\sin(\l_1+\xi-u-\eta) \sin(\l_2+\xi-u-\eta)}, \no\\
&& {k^+}^2_2(u)=\frac{2\cos(\l_1-\l_2)e^{2iu+i\eta}-
\cos(\l_1+\l_2+2\xi)e^{-i\eta}}
{4\sin(\l_1+\xi-u-\eta)\sin(\l_2+\xi-u-\eta)}.\label{K-matrix-6}
\eea The K-matrices depend on three free boundary parameters
$\{\l_1,\,\l_2,\,\xi\}$ which specify integrable boundary
conditions \cite{Gho94}. It is very convenient to introduce a
vector $\l=\sum_{k=1}^2\l_k\e_k$ associated with the boundary
parameters $\{\l_i\}$, where $\{\e_i,\,i=1,2\}$ form the
orthonormal basis of $V$ such that $\langle
\e_i,\e_j\rangle=\d_{ij}$. We remark that $K^-(u)$ does not depend
on the crossing parameter $\eta$ but $K^+(u)$ does. They satisfy
the following relation: \bea \lim_{\eta\rightarrow
0}\{K^+(u)K^-(u)\}=\lim_{\eta\rightarrow 0}\{K^+(u)\}K(u)={\rm
id}.\label{ID-1}\eea

Let us introduce the generalized  XXZ Gaudin operators
\cite{Gau76} $\{H_j |j=1,2,\cdots,N\}$ associated with the
spin-$\frac{1}{2}$ XXZ  model with generic boundaries specified by
the boundary K-matrices in (\ref{K-matrix}) and (\ref{DK-matrix}):
\bea &&H_j=\G_j(z_j)+\sum_{k\neq
j}^{2M}\frac{1}{\sin(z_j-z_k)}\lt\{\s^+_k\s^-_j+\s^-_k\s^+_j
+\cos(z_j-z_k)\frac{\s^z_k\s^z_j-1}{2}\rt\}\no\\
&&~~+\sum_{k\neq
j}^{2M}\frac{K_j^{-1}(z_j)}{\sin(z_j+z_k)}\lt\{\s^+_j\s^-_k+\s^-_j\s^+_k
+\cos(z_j+z_k)\frac{\s^z_j\s^z_k-1}{2}\rt\}K_j(z_j),\label{Ham}
\eea  where $\G_j(u)=\frac{\partial}{\partial
\eta}\{\bar{K}_j(u)\}|_{\eta=0}K_j(u)$, $j=1,\cdots,N,$ with
$\bar{K}_j(u)=tr_0\lt\{K^+_0(u)\R_{0j}(2u)P_{0j}\rt\}$, and
$\{z_j\}$ correspond to the inhomogeneous parameters of the
inhomogeneous spin-$\frac{1}{2}$ XXZ chain with generic open
boundary. For a generic choice of the boundary parameters
$\{\l_1,\,\l_2,\,\xi\}$, $\G_j(u)$ is an non-diagonal matrix, in
contrast to that of \cite{Lor02}.

The XXZ Gaudin operators (\ref{Ham}) are obtained by expanding the
double-row transfer matrix $\t(u)$ (\ref{trans}) at the point
$u=z_j$ around $\eta=0$: \bea &&\t(z_j)=\t(z_j)|_{\eta=0}+\eta
H_j+O(\eta^2),~~j=1,\cdots,N, \label{trans-2}\\
&&H_j=\frac{\partial}{\partial
\eta}\t(z_j)|_{\eta=0}.\label{Eq-1}\eea The relations
(\ref{quasi}) and (\ref{ID-1}) imply that the first term
$\t(z_j)|_{\eta=0}$ in the expansion (\ref{trans-2}) is equal to
an identity, namely, \bea \t(z_j)|_{\eta=0}={\rm id}.\label{First}
\eea Then the commutativity of the transfer matrices $\{\t(z_j)\}$
(\ref{Com-2}) for a generic $\eta$ implies \bea
[H_j,H_k]=0,~~i,j=1,\cdots,N.\label{Com-1}\eea Thus the Gaudin
system defined by (\ref{Ham}) is integrable. Moreover, the fact
that the Gaudin operators $\{H_j\}$ (\ref{Ham}) can be expressed
in terms of the transfer matrix of the inhomogeneous
spin-$\frac{1}{2}$ XXZ open chain enables us to exactly
diagonalize the operators by the algebraic Bethe ansatz method
with the help of the ``vertex-face" correspondence technique, as
can be seen in the next section. The aim of this paper is to find
the common eigenvectors and eigenvalues of the operators
(\ref{Ham}).

%%%%%%%%%%%%%%%%%%%%%%%%%%%%%%%%%%%%%%%%%%%%%%%%%%%%%%%%%%%%%%%
%                                                             %
%  4. Eigenvalues and Bethe ansatz equations                  %
%                                                             %
%                                                             %
%%%%%%%%%%%%%%%%%%%%%%%%%%%%%%%%%%%%%%%%%%%%%%%%%%%%%%%%%%%%%%%

\section{Eigenvalues and Bethe ansatz equations}
\label{BAE} \setcounter{equation}{0}
\subsection{Six-vertex SOS R-matrix and face-vertex
correspondence}

The simple root $\a$ and fundamental weight $\L_1$ of $sl(2)$ are
given in terms of the orthonormal basis  $\{\e_i\}$ as:
$\a=\e_1-\e_2,\,\L_1=\frac{\a}{2}$. Set \bea
\hat{i}=\e_i-\overline{\e},~~\overline{\e}=
\frac{1}{2}\sum_{k=1}^{2}\e_k,~~i=1,2,~~{\rm
then}~~\sum_{i=1}^2\hat{i}=0. \label{Vectors} \eea For each
dominant weight $\L=a\L_{1}~,~~a\in \Zb^+$(the set of non-negative
integer), there exists an irreducible highest weight
finite-dimensional representation $V_{\L}$ of $A_{1}$ with the
highest vector $ |\L\rangle$. For example the fundamental vector
representation is $V_{\L_1}$.

Let $\h$ be the Cartan subalgebra of $A_{1}$ and $\h^{*}$ be its
dual. A finite dimensional diagonalisable  $\h$-module is a
complex finite dimensional vector space $W$ with a weight
decomposition $W=\oplus_{\mu\in \h^*}W[\mu]$, so that $\h$ acts on
$W[\mu]$ by $x\,v=\mu(x)\,v$, $(x\in \h,~~v\in~W[\mu])$. For
example, the fundamental vector representation $V_{\L_1}=V$, the
non-zero weight spaces $W[\hat{i}]=\Cb \e_i,~i=1,2$.

For a generic $m\in V$, define \bea m_i=\langle m,\e_i\rangle,
~~m_{ij}=m_i-m_j=\langle m,\e_i-\e_j\rangle,~~i,j=1,2.
\label{Def1}\eea Let $R(u,m)\in {\rm End}(V\otimes V)$ be the
R-matrix of the six-vertex SOS model, which is trigonometric limit
of the eight-vertex SOS model \cite{Bax82} given by \bea
&&R(u,m)\hspace{-0.1cm}=\hspace{-0.1cm}
\sum_{i=1}^{2}R^{ii}_{ii}(u,m)E_{ii}\hspace{-0.1cm}\otimes\hspace{-0.1cm}
E_{ii}\hspace{-0.1cm}+\hspace{-0.1cm}\sum_{i\ne
j}^2\lt\{R^{ij}_{ij}(u,m)E_{ii}\hspace{-0.1cm}\otimes\hspace{-0.1cm}
E_{jj}\hspace{-0.1cm}+\hspace{-0.1cm}
R^{ji}_{ij}(u,m)E_{ji}\hspace{-0.1cm}\otimes\hspace{-0.1cm}
E_{ij}\rt\}, \label{R-matrix} \eea where $E_{ij}$ is the matrix
with elements $(E_{ij})^l_k=\d_{jk}\d_{il}$. The coefficient
functions are \bea &&R^{ii}_{ii}(u,\l)=1,~~
R^{ij}_{ij}(u,\l)=\frac{\sin u\sin(m_{ij}-\eta)}
{\sin(u+\eta)\sin(m_{ij})},~~i\neq j,\label{Elements1}\\
&& R^{ji}_{ij}(u,m)=\frac{\sin\eta\sin(u+m_{ij})}
{\sin(u+\eta)\sin(m_{ij})},~~i\neq j,\label{Elements2}\eea  and
$m_{ij}$ is defined in (\ref{Def1}). The R-matrix satisfies the
dynamical (modified) quantum Yang-Baxter equation \bea
&&R_{12}(u_1-u_2,m-\eta h^{(3)})R_{13}(u_1-u_3,m)
R_{23}(u_2-u_3,m-\eta h^{(1)})\no\\
&&~~~~=R_{23}(u_2-u_3,m)R_{13}(u_1-u_3,m-\eta
h^{(2)})R_{12}(u_1-u_2,m). \label{MYBE}\eea We adopt the notation:
$R_{12}(u,m-\eta h^{(3)})$ acts on a tensor $v_1\otimes v_2
\otimes v_3$ as $R(u,m-\eta\mu)\otimes id$ if $v_3\in W[\mu]$.

Define the following functions\bea \theta^{(1)}(u)=e^{-iu},~~
\theta^{(2)}(u)=1.\label{Fun}\eea Let us introduce an intertwiner,
i.e. an $2$-component  column vector $\phi_{m,m-\eta\hat{j}}(u)$
whose $k$-th element is \bea
\phi^{(k)}_{m,m-\eta\hat{j}}(u)=\theta^{(k)}(u+2m_j).\label{Intvect}\eea
Using the intertwiner, one can derive the following face-vertex
correspondence relation \cite{Cao03}\bea &&\R_{12}(u_1-u_2)
\phi_{m,m-\eta\hat{\imath}}(u_1)\otimes
\phi_{m-\eta\hat{\imath},m-\eta(\hat{\imath}+\hat{\jmath})}(u_2)
\no\\&&~~~~~~= \sum_{k,l}R(u_1-u_2,m)^{kl}_{ij}
\phi_{m-\eta\hat{l},\l-\eta(\hat{l}+\hat{k})}(u_1)\otimes
\phi_{m,m-\eta\hat{l}}(u_2). \label{Face-vertex}\eea Then the QYBE
(\ref{QYB}) of for the vertex-type R-matrix $\R(u)$ is equivalent
to the dynamical Yang-Baxter equation (\ref{MYBE}) of the SOS
R-matrix $R(u,m)$. For a generic $m$, we can introduce other types
of intertwiners $\bar{\phi},~\tilde{\phi}$ satisfying the
conditions, \bea
&&\sum_{k=1}^2\bar{\phi}^{(k)}_{m,m-\eta\hat{\mu}}(u)
~\phi^{(k)}_{m,m-\eta\hat{\nu}}(u)=\d_{\mu\nu},\label{Int1}\\
&&\sum_{k=1}^2\tilde{\phi}^{(k)}_{m+\eta\hat{\mu},m}(u)
~\phi^{(k)}_{m+\eta\hat{\nu},\l}(u)=\d_{\mu\nu},\label{Int2}\eea
from which one can derive the following relations:\bea
&&\sum_{\mu=1}^2\bar{\phi}^{(i)}_{m,m-\eta\hat{\mu}}(u)
~\phi^{(j)}_{m,m-\eta\hat{\mu}}(u)=\d_{ij},\label{Int3}\\
&&\sum_{\mu=1}^2\tilde{\phi}^{(i)}_{m+\eta\hat{\mu},m}(u)
~\phi^{(j)}_{m+\eta\hat{\mu},m}(u)=\d_{ij}.\label{Int4}\eea

Through straightforward calculations, we find the K-matrices
$K^{\pm}(u)$ given by (\ref{K-matrix}) and (\ref{DK-matrix}) can
be expressed in terms of the intertwiners and {\it diagonal\/}
matrices $\K(\l|u)$ and $\tilde{\K}(\l|u)$ as follows \bea
&&K^-(u)^s_t=
\sum_{i,j}\phi^{(s)}_{\l-\eta(\hat{\imath}-\hat{\jmath}),
~\l-\eta\hat{\imath}}(u)
\K(\l|u)^j_i\bar{\phi}^{(t)}_{\l,~\l-\eta\hat{\imath}}(-u),\label{K-F-1}\\
&&K^+(u)^s_t= \sum_{i,j}
\phi^{(s)}_{\l,~\l-\eta\hat{\jmath}}(-u)\tilde{\K}(\l|u)^j_i
\tilde{\phi}^{(t)}_{\l-\eta(\hat{\jmath}-\hat{\imath}),
~\l-\eta\hat{\jmath}}(u).\label{K-F-2}\eea Here the two {\it
diagonal\/} matrices $\K(\l|u)$ and $\tilde{\K}(\l|u)$ are given
by \bea &&\K(\l|u)\equiv{\rm Diag}(k(\l|u)_1,\,k(\l|u)_2)={\rm
Diag}(\frac{\sin(\l_1+\xi-u)}{\sin(\l_1+\xi+u)},\,
\frac{\sin(\l_2+\xi-u)}{\sin(\l_2+\xi+u)}),\label{K-F-3}\\
&&\tilde{\K}(\l|u)\equiv{\rm
Diag}(\tilde{k}(\l|u)_1,\,\tilde{k}(\l|u)_2)\no\\
&&~~~~~~~~~={\rm
Diag}(\frac{\sin(\l_{12}\hspace{-0.1cm}-\hspace{-0.1cm}
\eta)\sin(\l_1\hspace{-0.1cm}+\hspace{-0.1cm}\xi+\hspace{-0.1cm}u
\hspace{-0.1cm}+\hspace{-0.1cm}\eta)}
{\sin\l_{12}\sin(\l_1+\xi-u-\eta)},\,
\frac{\sin(\l_{12}\hspace{-0.1cm}+\hspace{-0.1cm}
\eta)\sin(\l_2\hspace{-0.1cm}+\hspace{-0.1cm}\xi\hspace{-0.1cm}
+\hspace{-0.1cm}u\hspace{-0.1cm}+\hspace{-0.1cm}\eta)}
{\sin\l_{12}\sin(\l_2+\xi-u-\eta)}).\label{K-F-4} \eea Moreover,
one can check that the matrices $\K(\l|u)$ and $\tilde{\K}(\l|u)$
satisfy the SOS type reflection equation and its dual,
respectively \cite{Yan03}. Although the K-matrices $K^{\pm}(u)$
given by (\ref{K-matrix}) and (\ref{DK-matrix}) are generally
non-diagonal (in the vertex form), after the face-vertex
transformations (\ref{K-F-1}) and (\ref{K-F-2}), the face type
counterparts $\K(\l|u)$ and $\tilde{\K}(\l|u)$ {\it
simultaneously\/} become diagonal. This fact enables us to apply
the generalized algebraic Bethe ansatz method developed in
\cite{Yan04} for SOS type integrable models  to diagonalize the
transfer matrices $\t(u)$ (\ref{trans}).

\subsection{Algebraic Bethe ansatz}

Using the relations (\ref{Int3}) and (\ref{Int4}), the
decomposition of $K^+(u)$ (\ref{K-F-2}) and the diagonal property
(\ref{K-F-4}), the transfer matrix $\t(u)$ (\ref{trans}) can be
recasted  into the following face type form   \bea
&&\t(u)=tr(K^+(u)\mathbb{T}(u))\no\\
&&~~=\sum_{\mu,\nu}tr\lt(K^+(u)
\phi_{\l-\eta(\hat{\mu}-\hat{\nu}),
\l-\eta\hat{\mu}}(u)\tilde{\phi}_{\l-\eta(\hat{\mu}-\hat{\nu}),
\l-\eta\hat{\mu}}(u)~\mathbb{T}(u) \phi_{\l, \l-\eta\hat{\mu}}(-u)
\bar{\phi}_{\l, \l-\eta\hat{\mu}}(-u)\rt)\no\\
&&~~=\sum_{\mu,\nu}\bar{\phi}_{\l, \l-\eta\hat{\mu}}(-u)K^+(u)
\phi_{\l-\eta(\hat{\mu}-\hat{\nu}),\l-\eta\hat{\mu}}(u)~
\tilde{\phi}_{\l-\eta(\hat{\mu}-\hat{\nu}),
\l-\eta\hat{\mu}}(u)~\mathbb{T}(u)\phi_{\l, \l-\eta\hat{\mu}}(-u)\no\\
&&~~=\sum_{\mu,\nu}\tilde{\K}(\l|u)_{\nu}^{\mu}\T(\l|u)^{\nu}_{\mu}=
\sum_{\mu}\tilde{k}(\l|u)_{\mu}\T(\l|u)^{\mu}_{\mu}.
\label{De1}\eea Here we have introduced the face-type double-row
monodromy matrix $\T(\l|u)$ \bea
&&\T(\l|u)^{\nu}_{\mu}=\tilde{\phi}_{\l-\eta(\hat{\mu}-\hat{\nu}),
\l-\eta\hat{\mu}}(u)~\mathbb{T}(u)\phi_{\l,
\l-\eta\hat{\mu}}(-u)\no\\
&&~~~~~~~~~~\equiv
\sum_{i,j}\tilde{\phi}^{(j)}_{\l-\eta(\hat{\mu}-\hat{\nu}),
\l-\eta\hat{\mu}}(u)~\mathbb{T}(u)^j_i\phi^{(i)}_{\l,
\l-\eta\hat{\mu}}(-u).\label{Mon-F} \eea This face-type double-row
monodromy matrix can  be expressed in terms of the face type
R-matrix $R(\l|u)$ (\ref{R-matrix}) and K-matrix $\K(\l|u)$
(\ref{K-F-3}) (for the details, see equation (4.19) of
\cite{Yan04}). Moreover  from (\ref{Relation-Re}),
(\ref{Face-vertex}) and (\ref{Int4}) one can derive the following
exchange relations among $\T(\l|u)^{\nu}_{\mu}$: \bea
&&\sum_{i_1,i_2}\sum_{j_1,j_2}~
R(u_1-u_2,\l)^{i_0,j_0}_{i_1,j_1}\T(\l+\eta(\hat{\jmath}_1+\hat{\imath}_2)|u_1)
^{i_1}_{i_2}\no\\
&&~~~~~~~~\times R(u_1+u_2,\l)^{j_1,i_2}_{j_2,i_3}
\T(\l+\eta(\hat{\jmath}_3+\hat{\imath}_3)|u_2)^{j_2}_{j_3}\no\\
&&~~=\sum_{i_1,i_2}\sum_{j_1,j_2}~
\T(\l+\eta(\hat{\jmath}_1+\hat{\imath}_0)|u_2)
^{j_0}_{j_1}R(u_1+u_2,\l)^{i_0,j_1}_{i_1,j_2}\no\\
&&~~~~~~~~\times\T(\l+\eta(\hat{\jmath}_2+\hat{\imath}_2)|u_1)^{i_1}_{i_2}
R(u_1-u_2,\l)^{j_2,i_2}_{j_3,i_3}.\label{RE-F} \eea

As in \cite{Yan04}, let us introduce a set of standard notions for
convenience: \footnote{The scalar factors in the definitions of
the operators $\B(\l|u)$ and $\D(\l|u)$ are to make the relevant
commutation relations as concise  as (\ref{Rel-1})-(\ref{Rel-3}).
} \bea &&\A(\l|u)=\T(\l|u)^1_1,~~\B(\l|u)=
\frac{\T(\l|u)^1_2}{\sin(\l_{12})},\label{Def-AB}\\
&&\D(\l|u)=\frac{\sin(\l_{12}+\eta)}{\sin(\l_{12})}
\{\T(\l|u)^2_2-R(2u,\l+\eta\hat{1})^{2\,1}_{1\,2}\A(\l|u)\}.
\label{Def-D}\eea Hereafter, we adopt the convention:
$\l_{ij}=\l_i-\l_j$, introduced in (\ref{Def1}). After  tedious
calculations, we find the commutation relations among $\A(\l|u)$,
$\D(\l|u)$ and $\B(\l|u)$. Here we give the relevant ones for our
purpose,\bea &&\A(\l|u)\B(\l-2\eta\hat{1}|v)=
\frac{\sin(u+v)\sin(u-v-\eta)}{\sin(u+v+\eta)\sin(u-v)}
\B(\l-2\eta\hat{1}|v)\A(\l-2\eta\hat{1}|u)\no\\
&&~~~~~~~~-\frac{\sin\eta\sin 2v}{\sin(u-v)\sin(2v+\eta)}
\frac{\sin(u-v-\l_{12}+\eta)} {\sin(\l_{12}-\eta)}
\B(\l-2\eta\hat{1}|u)\A(\l-2\eta\hat{1}|v)\no\\
&&~~~~~~~~-\frac{\sin\eta}{\sin(u+v+\eta)}
\frac{\sin(u+v+\l_{21}+2\eta)}{\sin(\l_{21}+\eta)}
\B(\l-2\eta\hat{1}|u)
\D(\l-2\eta\hat{1}|v),\label{Rel-1}\\
&&\D(\l|u)\B(\l-2\eta\hat{1}|v)=
\frac{\sin(u-v+\eta)\sin(u+v+2\eta)}{\sin(u-v)\sin(u+v+\eta)}
\B(\l-2\eta\hat{1}|v)\D(\l-2\eta\hat{1}|u)\no\\
&&~~~~~~~~-\frac{\sin\eta\sin(2u+2\eta)\sin(u-v+\l_{12}-\eta)}
{\sin(u-v)\sin(2u+\eta)\sin(\l_{12}-\eta)} \B(\l-2\eta\hat{1}|u)
\D(\l-2\eta\hat{1}|v)\no\\
&&~~~~~~~~+\frac{\sin\eta\sin 2v\sin(2u+2\eta)\sin(u-v+\l_{12})}
{\sin(u+v+\eta)\sin(2v+\eta)\sin(2u+\eta)\sin(\l_{12}-\eta)}\no\\
&&~~~~~~~~~~~~~~\times\B(\l-2\eta\hat{1}|u)
\A(\l-2\eta\hat{1}|v),\label{Rel-2}\\
&&\B(\l-2\eta\hat{1}|u)\B(\l-4\eta\hat{1}|v)
=\B(\l-2\eta\hat{1}|v) \B(\l-4\eta\hat{1}|u).\label{Rel-3}\eea
Here we have used the identity $\hat{2}=-\hat{1}$ which can be
derived from (\ref{Vectors}).

In order to apply the algebraic Bethe ansatz method, in addition
to the relevant commutation relations (\ref{Rel-1})-(\ref{Rel-3}),
one needs to construct a pseudo-vacuum state (also called
reference state) which is the common eigenstate of the operators
$\A$, $\D$ and is annihilated by the operator $\C$. In contrast to
the case of the spin-$\frac{1}{2}$ XXZ open chain with {\it
diagonal\/} $K^{\pm}(u)$ \cite{Skl88}, for the open chain with
generic {\it non-diagonal\/} K-matrices (\ref{K-matrix}) and
(\ref{DK-matrix}), the usually highest-weight state\bea
\lt(\begin{array}{l}1\\0\end{array}\rt)\otimes\cdots\otimes
\lt(\begin{array}{l}1\\0\end{array}\rt),\no\eea is no longer  the
pseudo-vacuum state. However, after the face-vertex
transformations (\ref{K-F-1}) and (\ref{K-F-2}), the face type
counterparts K-matrices $\K(\l|u)$ and $\tilde{\K}(\l|u)$ {\it
simultaneously\/} become diagonal. This suggests that one can
transfer the spin-$\frac{1}{2}$ XXZ open chain with generic
non-diagonal K-matrices into the corresponding SOS model with {\it
diagonal} K-matrices $\K(\l|u)$ and $\tilde{\K}(\l|u)$ given by
(\ref{K-F-3})-(\ref{K-F-4}) (possibly after some local gauge
transformations \cite{Cao03}). Then it is easy to construct the
pseudo-vacuum in the ``face language" and use the generalized
algebraic Bethe ansatz method \cite{Yan04} to diagonalize the
transfer matrix. We shall develop this game in the following.

Let us introduce the corresponding pseudo-vacuum state
$|\O\rangle$ \bea
|\O\rangle=\phi_{\l-(N-1)\eta\hat{1},\l-N\eta\hat{1}}(-z_1)\otimes
\phi_{\l-(N-2)\eta\hat{1},\l-(N-1)\eta\hat{1}}(-z_{2})\cdots\otimes
\phi_{\l,\l-\eta\hat{1}}(-z_N).\label{Vac} \eea The state does
only depend on the boundary parameters $\{\l_1,\l_2\}$ and the
inhomogeneous parameters $\{z_j\}$. Using the technique developed
in \cite{Yan04}, after tedious calculations, we find that the
pseudo-vacuum state given by (\ref{Vac}) satisfies the following
equations, as required, \bea
&&\A(\l-N\eta\hat{1}|u)|\O\rangle=\frac{\sin(\l_1+\xi-u)}{\sin(\l_1+\xi+u)}
|\O\rangle,\label{A}\\
&&\D(\l-N\eta\hat{1}|u)|\O\rangle=\frac{\sin
2u\sin(\l_1+\xi+u+\eta)\sin(\l_2+\xi-u-\eta)}
{\sin(2u+\eta)\sin(\l_1+\xi+u)\sin(\l_2+\xi+u)}\no\\
&&~~~~~~~~~~~~~~~~~~~~~~~~~~~~\times\lt\{\prod_{k=1}^N
\frac{\sin(u+z_k)\sin(u-z_k)}{\sin(u+z_k+\eta)\sin(u-z_k+\eta)}\rt\}
|\O\rangle,\label{D}\\
&&\C(\l-N\eta\hat{1}|u)|\O\rangle=0,\\
&&\B(\l-N\eta\hat{1}|u)|\O\rangle\neq 0. \eea Then  the so-called
Bethe states can be constructed by applying the creation operator
$\B$ on the pseudo-vacuum state \bea
|v_1,\cdots,v_M\rangle=\B(\l-2\eta\hat{1}|v_1)
\B(\l-4\eta\hat{1}|v_2)\cdots\B(\l-2M\eta\hat{1}|v_M)|\O\rangle.
\label{Bethe-state}\eea

From (\ref{De1}), (\ref{Def-AB}) and (\ref{Def-D}) we can rewrite
the transfer matrices $\t(u)$ (\ref{trans}) in terms of the
operators $\A$ and $\D$ \bea
&&\t(u)=\frac{\sin(\l_2+\xi-u)\sin(\l_1+\xi+u)\sin(2u+2\eta)}
{\sin(\l_2+\xi-u-\eta)\sin(\l_1+\xi-u-\eta)\sin(2u+\eta)}\A(\l|u)
\no\\
&&~~~~~~~~~~~~~~~~~~+\frac{\sin(\l_2+\xi+u+\eta)}
{\sin(\l_2+\xi-u-\eta)}\D(\l|u). \label{Exp-trans}\eea Acting the
above expression of the transfer matrices on the Bethe  states
$|v_1,\cdots,v_M\rangle$ (\ref{Bethe-state}) and repeatedly using
the relevant commutation relations (\ref{Rel-1})-(\ref{Rel-3}), we
obtain \bea
&&\hspace{-0.2cm}\t(u)|v_1,\cdots,v_M\rangle\hspace{-0.1cm}=\hspace{-0.1cm}
t(u)|v_1,\cdots,v_M\rangle\hspace{-0.1cm}+\hspace{-0.1cm}
\frac{\sin(2u+2\eta)\sin\eta}
{\sin(\l_2+\xi-u-\eta)\sin(\l_1+\xi-u-\eta)}\no\\
&&~~~~~~~~~~~~~~~~~~\times\lt\{\sum_{\a=1}^{M}
F_{\a}|v_1,\cdots,v_{\a-1},u,v_{\a+1},\cdots,v_M\rangle\rt\}.\label{Exch-1}\eea
Here, the term associated with function $t(u)$ is the so-called
{\it wanted\/} term which gives rise to the eigenvalues and the
term associated with $\{F_{\a}|\a=1,\cdots,M\}$ is the so-called
{\it unwanted\/} term. They are given, respectively, by \bea
&&t(u)=\frac{\sin(\l_2+\xi-u)\sin(\l_1+\xi-u)\sin(2u+2\eta)}
{\sin(\l_2+\xi-u-\eta)\sin(\l_1+\xi-u-\eta)\sin(2u+\eta)}\no\\
&&~~~~~~~~~~~~~~~~~~\times\prod_{k=1}^M\frac{\sin(u+v_k)\sin(u-v_k-\eta)}
{\sin(u+v_k+\eta)\sin(u-v_k)}\no\\
&&~~~~~~+\frac{\sin(\l_2+\xi+u+\eta)\sin(\l_1+\xi+u+\eta)\sin 2u}
{\sin(\l_2+\xi+u)\sin(\l_1+\xi+u)\sin(2u+\eta)}\no\\
&&~~~~~~~~~~~~~~~~~~\times\prod_{k=1}^M\frac{\sin(u+v_k+2\eta)\sin(u-v_k+\eta)}
{\sin(u+v_k+\eta)\sin(u-v_k)}\no\\
&&~~~~~~~~~~~~~~~~~~\times\prod_{k=1}^{2M}\frac{\sin(u+z_k)\sin(u-z_k)}
{\sin(u+z_k+\eta)\sin(u-z_k+\eta)},\label{Eigenfuction}
 \eea and
\bea &&F_{\a}=\frac{\sin
2v_{\a}\sin(\l_2+\xi-v_{\a})\sin(\l_1+\xi-v_{\a})}{\sin(2v_{\a}+\eta)
\sin(u+v_{\a}+\eta)\sin(u-v_{\a})}\no\\
&&~~~~~~\times\lt\{ \prod_{k\neq
\a}^M\frac{\sin(v_{\a}+v_k)\sin(v_{\a}-v_k-\eta)}
{\sin(v_{\a}+v_k+\eta)\sin(v_{\a}-v_k)}\rt.\no\\
&&~~~~~~~~-\hspace{-0.1cm}\frac{\sin(\l_2\hspace{-0.1cm}+\hspace{-0.1cm}
\xi\hspace{-0.1cm}+\hspace{-0.1cm}v_{\a}
\hspace{-0.1cm}+\hspace{-0.1cm}\eta)
\sin(\l_2\hspace{-0.1cm}+\hspace{-0.1cm}\xi\hspace{-0.1cm}-\hspace{-0.1cm}v_{\a}
\hspace{-0.1cm}-\hspace{-0.1cm}\eta)
\sin(\l_1\hspace{-0.1cm}+\hspace{-0.1cm}\xi\hspace{-0.1cm}+\hspace{-0.1cm}
v_{\a}\hspace{-0.1cm}+\hspace{-0.1cm}\eta)
\sin(\l_1\hspace{-0.1cm}+\hspace{-0.1cm}\xi\hspace{-0.1cm}-\hspace{-0.1cm}v_{\a}
\hspace{-0.1cm}-\hspace{-0.1cm}\eta)}
{\sin(\l_2+\xi+v_{\a})\sin(\l_2+\xi-v_{\a})
\sin(\l_1+\xi+v_{\a})\sin(\l_1+\xi-v_{\a})}\no\\
&&~~~~~~~~~~~~~~~~~~\times\prod_{k\neq
\a}^M\frac{\sin(v_{\a}+v_k+2\eta)\sin(v_{\a}-v_k+\eta)}
{\sin(v_{\a}+v_k+\eta)\sin(v_{\a}-v_k)}\no\\
&&~~~~~~~~~~~~~~~~~~\times\lt.\prod_{k=1}^{2M}\frac{\sin(v_{\a}+z_k)\sin(v_{\a}-z_k)}
{\sin(v_{\a}+z_k+\eta)\sin(v_{\a}-z_k+\eta)}\rt\},~~\a=1,\cdots,M.
\label{Unwanted}\eea The relation (\ref{Exch-1}) tells us that the
Bethe states $|v_1,\cdots,v_M\rangle$ are eigenstates of the
transfer matrices $\t(u)$ if the {\it unwanted\/} terms vanish:
$F_{\a}=0,\,\a=1,\cdots,M$. This leads to the Bethe ansatz
equations, \bea &&\hspace{-0.1cm}\frac
{\sin(\l_2+\xi+v_{\a})\sin(\l_2+\xi-v_{\a})
\sin(\l_1+\xi+v_{\a})\sin(\l_1+\xi-v_{\a})}
{\sin(\l_2\hspace{-0.1cm}+\hspace{-0.1cm}
\xi\hspace{-0.1cm}+\hspace{-0.1cm}v_{\a}
\hspace{-0.1cm}+\hspace{-0.1cm}\eta)
\sin(\l_2\hspace{-0.1cm}+\hspace{-0.1cm}\xi\hspace{-0.1cm}-\hspace{-0.1cm}v_{\a}
\hspace{-0.1cm}-\hspace{-0.1cm}\eta)
\sin(\l_1\hspace{-0.1cm}+\hspace{-0.1cm}\xi\hspace{-0.1cm}+\hspace{-0.1cm}
v_{\a}\hspace{-0.1cm}+\hspace{-0.1cm}\eta)
\sin(\l_1\hspace{-0.1cm}+\hspace{-0.1cm}\xi\hspace{-0.1cm}-\hspace{-0.1cm}v_{\a}
\hspace{-0.1cm}-\hspace{-0.1cm}\eta)}\no\\
&&~~~~~~=\prod_{k\neq
\a}^M\frac{\sin(v_{\a}+v_k+2\eta)\sin(v_{\a}-v_k+\eta)}
{\sin(v_{\a}+v_k)\sin(v_{\a}-v_k-\eta)}\no\\
&&~~~~~~~~~~\times\prod_{k=1}^{2M}\frac{\sin(v_{\a}+z_k)\sin(v_{\a}-z_k)}
{\sin(v_{\a}+z_k+\eta)\sin(v_{\a}-z_k+\eta)},~~\a=1,\cdots,M.
\label{BA}\eea We have checked that  the Bethe ansatz equations
indeed  ensure that the eigenvalues (\ref{Eigenfuction}) of
transfer matrices $\t(u)$ are entire functions. Our result
recovers that of \cite{Cao03} for the very special case
$z_k=0,\,k=1,\cdots,2M$.

\subsection{Eigenstates and the corresponding eigenvalues}
The relation (\ref{Eq-1}) between $\{H_j\}$ and $\{\t(z_j)\}$ and
the fact that the first term of (\ref{trans-2}) is a c-number
enable us to extract the eigenstates of the Gaudin operators and
the corresponding eigenvalues from the results obtained in last
subsection.

Let us introduce the state $|\O^{(0)}\rangle$,\bea
|\O^{(0)}\rangle=\lt(\begin{array}{c}e^{i(z_1-2\l_1)}\\1\end{array}\rt)
\otimes\cdots\otimes
\lt(\begin{array}{c}e^{i(z_N-2\l_1)}\\1\end{array}\rt)\eea This
state can be obtained from the pseudo-vacuum state $|\O\rangle$
(\ref{Vac}) by taking the limit:
$|\O^{(0)}\rangle=\lim_{\eta\rightarrow 0}|\O\rangle$. Let us
introduce the matrix $C(u)\in {\rm End}(V)$ associated the
intertwiner vector $\phi$ \bea
C(u)=\lt(\begin{array}{cc}e^{-i(u+2\l_1)}&e^{-i(u+2\l_2)}
\\1&1\end{array}\rt),\label{Matrix-in}\eea
and the associated gauged Pauli operator $\s^-(u)\in {\rm
End}(V)$, \bea \s^-(u)=C(u)\s^-C(u)^{-1}.\label{Matrix-in-1}\eea
Then we define the states $\Psi(v_1,\cdots,v_M)$ and $\Phi_{\a}$:
\bea &&\Psi(v_1,\cdots,v_M)=\prod_{\a=1}^M\lt(\sum_{k=1}^{2M}\lt\{
\frac{\sin(\l_1+\xi-v_{\a})\sin(v_{\a}-z_k+\l_{12})}
{\sin(\l_1+\xi+v_{\a})\sin(v_{\a}-z_k)}\rt.\rt.\no\\
&&~~~~~~~~~~~~~~~~~~~~-
\lt.\lt.\frac{\sin(\l_2+\xi-v_{\a})\sin(v_{\a}+z_k-\l_{12})}
{\sin(\l_2+\xi+v_{\a})\sin(v_{\a}+z_k)} \rt\}\s_k^-(-z_k)\rt)
|\O^{(0)}\rangle,\label{BAS-1}\\
&&\Phi_{\a}=\prod_{\b\neq \a}^M\lt(\sum_{k=1}^{2M}\lt\{
\frac{\sin(\l_1+\xi-v_{\b})\sin(v_{\b}-z_k+\l_{12})}
{\sin(\l_1+\xi+v_{\b})\sin(v_{\b}-z_k)}\rt.\rt.\no\\
&&~~~~~~~~~~~~~~~~~~~~-
\lt.\lt.\frac{\sin(\l_2+\xi-v_{\b})\sin(v_{\b}+z_k-\l_{12})}
{\sin(\l_2+\xi+v_{\b})\sin(v_{\b}+z_k)} \rt\}\s_k^-(-z_k)\rt)
|\O^{(0)}\rangle.\label{BAS-2}\eea

Noting the relations (\ref{Eq-1}), (\ref{trans-2}) and
(\ref{Exch-1})-(\ref{Unwanted}), and using the same method as in
\cite{Hik95}, we derive the following relations \bea
&&H_j\Psi(v_1,\cdots,v_M)=E_j\Psi(v_1,\cdots,v_M)+\sum_{\a=1}^M\lt\{\frac{\sin
2z_j\sin 2v_{\a}} {\sin(z_j+v_{\a})\sin(z_j- v_{\a})}\rt.\no\\
&&~~~~~~~~~~~~~~~~~~~~~~~~~~~\times
\lt.\frac{\sin(\l_2+\xi-v_{\a})\sin(\l_1+\xi-v_{\a})}
{\sin(\l_2+\xi-z_j)\sin(\l_1+\xi+z_j)}
f_{\a}\s_j^-(-z_j)\Phi_{\a}\rt\}.\label{Relation-BA}\eea The
functions $E_j$ and $f_{\a}$ are \bea &&E_j=\cot
2z_j+\sum_{j=1}^2\cot(\l_j+\xi-z_j)+\sum_{k=1}^M
\frac{\sin 2z_j}{\sin(v_k-z_j)\sin(v_k+z_j)},\label{Eig-1}\\
&&f_{\a}=\sum_{j=1}^2\frac{1}{\sin(\l_j+\xi-v_{\a})
\sin(\l_j+\xi+v_{\a})}+\sum_{k=1}^{2M} \frac{1}{\sin(v_{\a}+z_k)
\sin(v_{\a}-z_k)}\no\\
&&~~~~~~~~~~~~-2\sum_{k\neq \a}^M \frac{1}{\sin(v_{\a}+v_k)
\sin(v_{\a}-v_k)}.\eea The equation (\ref{Relation-BA}) suggests
that the state $\Psi(v_1,\cdots,v_M)$ is an eigenstate of the
Gaudin operators $\{H_j\}$ if $\{v_j|j=1,\cdots,M\}$ is set to
satisfy $f_{\a}=0$ for $\a=1,\cdots,M$. This leads to the
corresponding Bethe ansatz equations \bea
&&\sum_{j=1}^2\frac{1}{\sin(\l_j+\xi-v_{\a})
\sin(\l_j+\xi+v_{\a})}+\sum_{k=1}^{2M} \frac{1}{\sin(v_{\a}+z_k)
\sin(v_{\a}-z_k)}\no\\
&&~~~~~~~~~~~~=2\sum_{k\neq \a}^M \frac{1}{\sin(v_{\a}+v_k)
\sin(v_{\a}-v_k)},~~\a=1,\cdots,M.\label{BAE-1}\eea
\section{Conclusions}
\label{Con} \setcounter{equation}{0}

We have studied the XXZ Gaudin model with generic boundaries
specified by the non-diagonal K-matrices $K^{\pm}(u)$,
(\ref{K-matrix}) and (\ref{DK-matrix}). In addition to the
inhomogeneous parameters $\{z_j\}$, the associated Gaudin
operators $\{H_j\}$, (\ref{Ham}), have three free parameters
$\{\l_1,\l_2,\xi\}$, which  give rise to three-parameter
($\l_1,\l_2,\xi$) generalizations of those in \cite{Ami01} and
two-parameter ($\l_1,\l_2$) generalizations of those in
\cite{Hik95,Lor02}. As seen from section 4, although the ``vertex
type" K-matrices $K^{\pm}(u)$ (\ref{K-matrix}) and
(\ref{DK-matrix}) are {\it non-diagonal\/}, the compositions,
(\ref{K-F-1}) and (\ref{K-F-2}), lead to the {\it diagonal\/}
``face-type" K-matrices, (\ref{K-F-3}) and (\ref{K-F-4}), after
the face-vertex transformation. This enables us to successfuly
construct the corresponding pseudo-vacuum state $|\O\rangle$
(\ref{Vac}) and apply the algebraic Bethe ansatz method developed
in \cite{Yan04} for the SOS type models to diagonalize the
transfer matrix $\t(u)$ of the inhomogeneous spin-$\frac{1}{2}$
XXZ open chain with the generic non-diagonal K-matrices.
Furthermore, we have exactly diagonalized the Gaudin operators
$\{H_j\}$, and derived their common eigenstates (\ref{BAS-1}) and
eigenvalues (\ref{Eig-1}) as well as the associated Bethe ansatz
equations (\ref{BAE-1}).

One may choose the K-matrix $K^+(u)$ whose matrix elements are
given by \bea && {k^+}^1_1(u)= \frac{2\cos(\l_1-\l_2)e^{-i\eta}
-\cos(\l_1+\l_2+2\bar{\xi})e^{2iu+i\eta}}
{4\sin(\l_1+\bar{\xi}-u-\eta)
\sin(\l_2+\bar{\xi}-u-\eta)},\no\\
&&{k^+}^1_2(u)=\frac{i\sin(2u+2\eta)e^{-i(\l_1+\l_2)}
e^{iu-i\eta}} {2\sin(\l_1+\bar{\xi}-u-\eta)
\sin(\l_2+\bar{\xi}-u-\eta)},
\no\\
&&{k^+}^2_1(u)=\frac{-i\sin(2u+2\eta)e^{i(\l_1+\l_2)}
e^{iu+i\eta}}
{2\sin(\l_1+\bar{\xi}-u-\eta) \sin(\l_2+\bar{\xi}-u-\eta)}, \no\\
&& {k^+}^2_2(u)=\frac{2\cos(\l_1-\l_2)e^{2iu+i\eta}-
\cos(\l_1+\l_2+2\bar{\xi})e^{-i\eta}}
{4\sin(\l_1+\bar{\xi}-u-\eta)\sin(\l_2+\bar{\xi}-u-\eta)}.\label{K-matrix-D1}
\eea Taking $\bar{\xi}=\xi$, the above K-matrix recovers that
given by (\ref{K-matrix-6}). We have used the same method in
subsection 4.2 to diagonalize the transfer matrix $\tau(u)$
(\ref{trans}) with the non-diagonal K-matrices $K^-(u)$
(\ref{K-matrix-2-1}) and $K^+(u)$ (\ref{K-matrix-D1}). The
corresponding eigenvalue $t(u)$ is given by \bea
&&t(u)=\frac{\sin(\l_2+\bar\xi-u)\sin(\l_1+\bar\xi+u)\sin(\l_1+\xi-u)\sin(2u+2\eta)}
{\sin(\l_2+\bar\xi-u-\eta)\sin(\l_1+\bar\xi-u-\eta)\sin(\l_1+\xi+u)\sin(2u+\eta)}\no\\
&&~~~~~~~~~~~~~~~~~~\times\prod_{k=1}^M\frac{\sin(u+v_k)\sin(u-v_k-\eta)}
{\sin(u+v_k+\eta)\sin(u-v_k)}\no\\
&&~~~~~~+\frac{\sin(\l_2+\bar\xi+u+\eta)\sin(\l_1+\xi+u+\eta)\sin(\l_2+\xi-u-\eta)\sin
2u}
{\sin(\l_2+\bar\xi-u-\eta)\sin(\l_1+\xi+u)\sin(\l_2+\xi+u)\sin(2u+\eta)}\no\\
&&~~~~~~~~~~~~~~~~~~\times\prod_{k=1}^M\frac{\sin(u+v_k+2\eta)\sin(u-v_k+\eta)}
{\sin(u+v_k+\eta)\sin(u-v_k)}\no\\
&&~~~~~~~~~~~~~~~~~~\times\prod_{k=1}^{2M}\frac{\sin(u+z_k)\sin(u-z_k)}
{\sin(u+z_k+\eta)\sin(u-z_k+\eta)},\label{Eigenfuction-D-1}
 \eea  and the associated Bethe ansatz
equations are given by \bea &&\hspace{-0.1cm}\frac
{\sin(\l_2+\xi+v_{\a})\sin(\l_2+\bar\xi-v_{\a})
\sin(\l_1+\bar\xi+v_{\a})\sin(\l_1+\xi-v_{\a})}
{\sin(\l_2\hspace{-0.1cm}+\hspace{-0.1cm}
\bar\xi\hspace{-0.1cm}+\hspace{-0.1cm}v_{\a}
\hspace{-0.1cm}+\hspace{-0.1cm}\eta)
\sin(\l_2\hspace{-0.1cm}+\hspace{-0.1cm}\xi\hspace{-0.1cm}-\hspace{-0.1cm}v_{\a}
\hspace{-0.1cm}-\hspace{-0.1cm}\eta)
\sin(\l_1\hspace{-0.1cm}+\hspace{-0.1cm}\xi\hspace{-0.1cm}+\hspace{-0.1cm}
v_{\a}\hspace{-0.1cm}+\hspace{-0.1cm}\eta)
\sin(\l_1\hspace{-0.1cm}+\hspace{-0.1cm}\bar\xi\hspace{-0.1cm}-\hspace{-0.1cm}v_{\a}
\hspace{-0.1cm}-\hspace{-0.1cm}\eta)}\no\\
&&~~~~~~=\prod_{k\neq
\a}^M\frac{\sin(v_{\a}+v_k+2\eta)\sin(v_{\a}-v_k+\eta)}
{\sin(v_{\a}+v_k)\sin(v_{\a}-v_k-\eta)}\no\\
&&~~~~~~~~~~\times\prod_{k=1}^{2M}\frac{\sin(v_{\a}+z_k)\sin(v_{\a}-z_k)}
{\sin(v_{\a}+z_k+\eta)\sin(v_{\a}-z_k+\eta)},~~\a=1,\cdots,M.
\label{BA-D-1}\eea For the $\bar\xi=\xi$ case,
(\ref{Eigenfuction-D-1}) and (\ref{BA-D-1}) result in
(\ref{Eigenfuction}) and (\ref{BA}) respectively.

%%%%%%%%%%%%%%%%%%%%%%%%%%%%%%%%%%%%%%%%%%%%%%%%%%%%%%%%%%%%%%%
%                                                             %
%  Acknowledgments                                            %
%                                                             %
%%%%%%%%%%%%%%%%%%%%%%%%%%%%%%%%%%%%%%%%%%%%%%%%%%%%%%%%%%%%%%%
\section*{Acknowledgements}
The financial support from  Australian Research Council is
gratefully acknowledged.

%%%%%%%%%%%%%%%%%%%%%%%%%%%%%%%%%%%%%%%%%%%%%%%%%%%%%%%%%%%%%%%
%                                                             %
%  References                                                 %
%                                                             %
%%%%%%%%%%%%%%%%%%%%%%%%%%%%%%%%%%%%%%%%%%%%%%%%%%%%%%%%%%%%%%%

\end{document}